\documentstyle[preprint,aps,eqsecnum]{revtex}
\begin{document}
\title{Nonuniversality in level dynamics}
\author{Pawe{\l} Kunstman,  Karol
\.Zyczkowski, and Jakub Zakrzewski}
\address{
Instytut Fizyki Mariana Smoluchowskiego, Uniwersytet  Jagiello\'nski,
\\ ul. Reymonta 4,  30-059 Krak\'ow, Poland}
\date{\today}
\maketitle
\begin{abstract}

Statistical  properties of parametric motion in ensembles of
Hermitian banded random matrices are studied.
 We analyze the distribution of level velocities and
level curvatures as well as  their correlation functions
in the crossover regime between three universality classes.
 It is shown that the statistical properties of level
dynamics are in  general  {\it non} universal and strongly
depend on the way in which the parametric dynamics is introduced.

\end{abstract}

\newpage

\section{Introduction}

A link between random matrix theory (RMT) \cite{mehta}
and the statistical properties of spectra of quantum
systems is well established.
Depending on the symmetry of a classically chaotic
quantum system, its spectral fluctuations are
described by Gaussian orthogonal (GOE), Gaussian unitary (GUE) or
Gaussian symplectic ensemble (GSE) \cite{Ha91,Bo91}.

Quite often the physical systems depend on some external parameter,
say $\lambda$, therefore
it is  interesting to study the level dynamics i.e. the motion of
eigenvalues $E_i(\lambda)$ as a function of $\lambda$.
Among the first parametric properties studied were the investigations
of the avoided crossings gaps \cite{Wi89,GS91,ZK91}, the parametric
number variance \cite{GSBSWZ} or the curvature of the levels
(i.e. the
second derivatives of their energies with respect to the parameter)
\cite{Ga,SHG91,TH92,ZD93,Vo94,FS95}.
It has been claimed that
 the statistical properties of level dynamics are universal
for disordered or strongly chaotic systems
\cite{Ga,Alt} provided the change of $\lambda$ does not modify
global symmetries properties. To reveal the universality one
has both to unfold the energy levels \cite{Bo91} and appropriately
rescale the parameter, $\lambda$ \cite{Ga,ZD93,Alt}.
Other statistical
measures of parametric dynamics such as the level slopes (velocities)
distribution (Gaussian shaped for random systems \cite{Ga,ZD93,Alt}),
the
velocity-velocity correlation
function  \cite{Alt,Be93,Za95,GZZ95} in the bound spectrum or
parametric conductance fluctuations \cite{BLM96} and fluctuations
in the Wigner time delay \cite{SZZ96} for scattering systems have
also been discussed.

A word of caution is, however, necessary at this point. Even for
the nearest neighbour spacing distribution, widely considered to be
universal, exceptions from the RMT prediction may be quite
significant for real physical systems \cite{ZDD95}. Much more
pronounced and common are the deviations from the RMT predictions
for the parametric motion of levels. In particular, as shown by
Takami and Hasegawa \cite{TH92}, the curvature distribution
shows nonuniversal behaviour for small curvatures even for
the mixing system (a Bunimovich stadium). Similarly non Gaussian
slopes distribution as well as strongly non Cauchy-like curvature
distribution was observed for magnetized hydrogen atom \cite{ZD93}.
The origin of these deviations has been linked to partial
wavefunction localization on unstable periodic orbits. Thus
the non-generic features of parametric statistics may provide
most interesting information about the physics of a given
physical system.

Most of these studies considered pure symmetry cases, i.e.,
systems pertaining to a given, e.g., GOE universality class.
This is often not the case in a realistic situation. In particular
the presence of the magnetic field, or the Aharonov-Bohm flux
the time-reversal invariance symmetry (TRI) becomes broken --
such a situation corresponds to a crossover between GOE and GUE
for random system. In this context the velocity--velocity
correlation function has been studied intensively
\cite{SA93,BK94,GZZ95} as well as the velocity distribution
\cite{THSA} or the curvature distribution \cite{KB94,BM94}. The
authors considered mostly the situation when the increase of the
external parameter, $\lambda$ (e.g. the magnetic field) destroys
the time-reversal invariance although, importantly, it has been
noticed \cite{THSA} that the parametric velocity distribution
may strongly depend on the nature of the perturbation.
Relatively less frequent were studies of the parametric dynamics
in the transition region between completely delocalized and
localized spectrum (see, however, e.g., the treatment of the
velocity distribution for broken-TRI case in \cite{Fy94,FM95}).

In order to model the spectra of quantum systems in a crossover
regime (a weak localization or a partially broken symmetry)
one may utilize  random matrix ensembles that interpolate between
the canonical ensembles. For example,
real symmetric band random matrices are capable to  model
transition between the localized and the delocalized regime.
Statistical properties of their eigenvalues and eigenvectors
depend on a single scaling parameter $x=b^2/N$ \cite{CMI90},
where $N$ denotes the matrix size and $b$ the band width.
Allowing the matrices to be Hermitian and
changing the relative weight
of the imaginary component $\alpha$
 one can model the effect of the time-reversal
symmetry breaking and the transition from an orthogonal to an unitary
universality class. The corresponding
scaling parameter $y$ is  proportional to $N\alpha$
\cite{PM83} for small perturbations.
 An ensemble of Hermitian band random matrices (HBRM) can
be therefore completely characterized by two scaling parameters
$(x,y)$
\cite{ZS96}. The Poissonian, strongly localized spectrum is obtained
 in the limit $x <<1$,
 while in the opposite delocalized limit $x>>1$ the model reduces to
GOE for $y=0$ and GUE for $y>>1$.

This work is intended as a systematic study of
the parametric dynamics and the corresponding
statistical measures for the transition region and in the localized
regime. Our work differs from most of the analysis mentioned above
in a way a parametric dependence is introduced. We assume
that the changes of
$\lambda$ leave the global properties of the system unaffected.
In the random matrix approach this is equivalent to the assumption
that the statistical properties of the ensemble of matrices do not
depend on the value of the parameter, $\lambda$, determining  the parametric
dynamics. For the physical system applications this is equivalent to
saying that the symmetry properties of the system considered, but also
the character of the underlying classical dynamics (say, the fraction
of the phase space volume which is chaotic) are invariant with respect
to $\lambda$. Or, from a practical point of view,
the the dynamics changes only weakly with $\lambda$, in the interval of
$\lambda$ values considered in each case.
The level dynamics is, in a sense, "perpendicular"
to crossovers between canonical ensembles, as schematically shown in
Fig.~\ref{scheme}.
 Such a physical situation may
correspond to a variation of the disorder parameter in a mesoscopic
system, for which all other parameters are kept constant.
 A special attention is drawn to the localized case,
characterized by small values of $x$ where some analytic predictions
obtained using supersymmetric calculus exist \cite{Fy94,FM95}.

The interest in such a study is twofold. Firstly, it is interesting
to see how the parametric properties of the system follow the
transition between different pure universality classes. Secondly,
the results obtained in the HBRM model may serve in future as
a reference for a comparison with statistics obtained in
 real physical systems. It is
then of utmost importance to know what one may expect from the
purely random model. This may enable to isolate the non generic,
i.e. characteristic for a given system, properties.

 The paper is organized as follows. In the next section we
describe the model and the parametric dynamics.
The distribution of level velocities $P(v)$
is analyzed in Section III. The next
Section is devoted to velocity-velocity correlation function
$C_v(\lambda)$. Level curvatures are discussed is Sec. V, whereas their
correlations are investigated in Section VI. Next Sections
consider higher order statistical measures. Finally we discuss
the consequences of the results obtained for various
statistical measures in the concluding section.

\section{Parametric dynamics for Hermitian band random matrices}
\label{s:model}

Hermitian band random matrices are defined by
\begin{equation}
H_{ij}=(\xi^R_{ij}+i\xi^I_{ij})\Theta(b-|i-j|)\quad
i\mbox{,}j=1,\ldots N,
\label{matrix1}
\end{equation}
where $\Theta(.)$
 denotes the unit step function vanishing at the origin.
Independent random variables
$\xi^R_{ij}$ and $\xi^I_{ij}$
are distributed according to
Gaussian distributions with zero mean the root mean squares  equal  to
$\sigma_{ij}^{R}$ and $\sigma_{ij}^{I}$,
respectively.  The parameter $\alpha$ measures the relative
size of the imaginary part of the off diagonal matrix elements
$\alpha  = (\sigma_{ij}^I/\sigma^R_{ij})^2$,
$ \quad i \ne j$; (notation has been simplified
with respect to Ref. \cite{ZS96}).
A normalization condition, ${\rm Tr}\, (H^2)=N+1$,  keeps
all the eigenvalues in a constrained energy range.
It also
allows us to express the
variances of real and imaginary parts of matrix elements
in terms of matrix size $N$, integer band width $b$ and real
parameter $\alpha$
\begin{equation}
(\sigma_{ij}^R)^2={(N+1)\over{2N+(\alpha+1)(2N-b)(b-1)}}(1+\delta_{ij}),
\label{var1}
\end{equation}
\begin{equation}
(\sigma_{ij}^I)^2=
{\alpha(N+1)\over{2N+(\alpha+1)(2N-b)(b-1)}}(1-\delta_{ij}).
\label{var2}
\end{equation}

For a diagonal random matrix $(b=1)$ the density
of eigenvalues is Gaussian and the level spacings are distributed
according to the Poisson distribution,
independently of the parameter $\alpha$.
In the opposite limiting case
of the full matrix $(b=N)$ variations of the parameter $\alpha$
correspond to the process of the time reversal symmetry breaking
in a dynamical system and
control the transition between orthogonal ($\alpha=0$) and unitary
($\alpha=1$) ensembles. \par

Statistical properties of spectrum and eigenvectors of real symmetric
band matrices depend only on a
scaling parameter $x=b^2/N$.
This scaling law, observed first by numerical computation of
localization length \cite{CMI90}, was reported
to describe also the distribution of eigenvalues \cite{CIM91}
and eigenvectors \cite{ZLKI91}, and subsequently
 explained theoretically \cite{FM91}.

 The same scaling law
holds also for Hermitian matrices \cite{FM91,Zy93}.
Moreover, effects of the time-reversal symmetry breaking are
controlled by another scaling parameter
$y=2N\alpha/(1-\alpha)$ \cite{ZS96}, stemming from
the universal properties of orthogonal--unitary transition founded by
Pandey and Mehta \cite{PM83}. The structure of eigenfunctions of HBRM
and the distribution of inverse participation ratio has also been
studied recently \cite{IMZ96}.

Let us now consider the parametric random matrix
\begin{equation}
H(\lambda)=H_1 \cos\lambda+H_2 \sin\lambda.
\label{pardyn}
\end{equation}
Both matrices $H_1$ and $H_2$ are taken from the same ensemble of
HBRM.
 Hence the  spectral properties of $H$ are stationary and do not depend
on $\lambda$. Moreover, during the transition, the motion of
eigenvalues is restricted to a bounded energy interval for arbitrary $\lambda$.
 This model of parametric dynamics was already used
for  investigation of level curvatures \cite{ZD93} and
velocity correlation functions \cite{Za95,GZZ95}.
The dynamics of eigenvalues as a function of $\lambda$
may be treated as the dynamics of interacting particles
(eigenvalues) with $\lambda$ playing the rule of the
fictitious time \cite{Ha91}. This allows to interpret the
slope of the levels as the velocity of the particles and their
curvature as the corresponding accelerations.

Parametric dynamics defined above  can  be
 studied numerically is a straightforward way.
  For several values of ensemble parameters
 $(N,b,\alpha)$ we have generated random matrices according to
Eqs.~(\ref{matrix1}-\ref{pardyn}). Diagonalizations of
resulting matrices for several values of $\lambda$ have allowed us then
 to find level velocities and  curvatures by a finite difference method.
 A special care has been paid to obtain reliable values of
velocities and curvatures, especially for very small and very
large values, by varying the size of the step in $\lambda$
 \cite{Ku95}.
 Before computing the derivatives of eigenvalues with respect to
$\lambda$ the standard unfolding technique was applied \cite{Bo91}
to set the mean level spacing $\Delta$  to unity.
We have considered matrices of size $N$ varying between $50$ to $500$,
 velocities and curvatures have been computed at about
 $200$ different values of $\lambda$, and the
 typical number of the independent realizations of dynamics,
Eq.~(\ref{pardyn}), in each case studied has varied
with matrix size to ensure at least $200\ 000$
data in each statistics. In other words, we have
simultaneously performed the averaging over the energy
(data from different energy levels of a given matrix $H$)
 and the averaging
over the disorder parameter
(several realizations of the dynamics for the same values of $N,b$ and
$\alpha$).

To check reliability of the numerical procedure  we have
 set $b=N$ and we have  reproduced
known results concerning the distribution of velocities and curvatures
as well as the velocity correlation function for
GOE ($\alpha=0$) and GUE
($\alpha=1$).  Moreover,  we have verified that
both  scaling parameters, $x$ and $y$, correctly describe the
parametric dynamics. The statistical properties of all quantities
studied have been found to be independent of the matrix dimension $N$
(for sufficiently large $N$) provided
  the parameters
$x$ and $y$ have been kept constant.

In the following sections we describe novel
results obtained for different statistics,  commencing
 with the distribution of first derivatives,
i.e. velocities. To avoid any misunderstanding let us repeat again
 that all the data presented are
obtained for ``perpendicular'' transitions [both $H_1$ and $H_2$ in
Eq.~(\ref{pardyn}) belong to the {\it same} random matrix ensemble]
as exemplified by double-sided arrows in Fig.~\ref{scheme}.
Thus for all values of $\lambda$ the scaling parameters $x$ and $y$
have the same values. We shall not consider here the case when
the parameter change modifies the global symmetry properties
-- a situation exemplified by broken line arrow in Fig.~\ref{scheme}.

\section{Distribution of level velocities}

For level dynamics within GOE or GUE the
distribution of level velocities, $P(v)$, is Gaussian \cite{Ga,ZD93,Alt}.
This fact is easy to explain using the first order perturbation
theory. For $\lambda=0$ the derivative $dE_i/d\lambda$ is
equal
to the diagonal element of matrix $H_2$ expanded in the eigenbasis of
$H_1$. Since both matrices are drawn independently from the same
ensemble, the matrix elements are Gaussian random numbers leading to
 Gaussian velocity
distribution.

On the other hand, in  strongly localized limit an
analytical formula for $P(v)$ given by Fyodorov \cite{Fy94} for
systems with a broken TRI strongly
differs from a Gaussian.
 A non-Gaussian character of the velocity distribution
for the GOE $\rightarrow$ GUE transition,
corresponding to the TRI symmetry breaking, has been
discussed in \cite{THSA,KB94,BM94}.

We have analyzed the transition between localized and delocalized
spectra both for random systems with a broken TRI
(i.e., the ensembles interpolating between Poisson and GUE) and for
ensembles interpolating between Poisson and GOE. The former allows us
to test the analytical prediction of Fyodorov \cite{Fy94}.

The theoretical prediction, as presented in \cite{Fy94}, has no free
parameters, both the shape of the distribution and its scale
(determined by the velocity variance) are determined by the theory.
Surprisingly the direct comparison of that distribution with
the numerical data
obtained has been highly unsatisfactory. The agreement is recovered,
see Fig.~\ref{figu2},
when both the theoretical distribution and the numerically obtained
data are rescaled with respect to the velocity variance,
 $\sigma_v=\sqrt{\langle v^2 \rangle}$
(note that the mean velocity vanishes by the construction of the
ensemble). Thus the apparent disagreement originally observed is
due to the difference between the theoretical and numerically obtained
values of the velocity variance (the ratio of the numerical value to
the theoretical prediction being about 13).
 We do not have a clear
explanation of this disagreement. It may be due to the fact that
the band width in our HBRM ensemble is sharply defined
 [compare Eq.~(\ref{matrix1})] while Fyodorov \cite{Fy94} assumed a smooth
decrease of the random matrix elements variance with increasing
distance from the diagonal, $|i-j|$.

 The theoretical prediction \cite{Fy94}, represented by
a smooth line in Fig.~\ref{figu2},
 takes the form
\begin{equation}
\label{fyod}
P(w) =  {\pi\over 6} {\pi w {\rm{coth}}(\pi w/\sqrt{6}) -\sqrt{6} \over
{\rm{sinh}}^2(\pi w/\sqrt{6})
},
\end{equation}
where the rescaled velocity $w=v/\sigma_v$. The similar quality
agreement is obtained for different values of the scaling parameter
up to $x$ of the order of unity corresponding to the transition to
a delocalized case. Then the numerical data start to show Gaussian
(typical for GUE) large velocity tail instead of the exponential tail
corresponding to fully localized situation, as exemplified by the
lack of large velocities in the numerical data presented in
Fig.~\ref{figu2} as a thin line histogram.

Although the theoretical prediction is obtained for the case of
a fully broken TRI our numerical data indicate that it works
extremely well also for preserved TRI (real symmetric matrices)
provided that again the velocity variance is appropriately
adjusted. The exemplary data are presented in
Fig.~\ref{figu3}
 for two cases corresponding to strong localization and
a transition to the delocalized regime. Here the numerically obtained
variance is twice larger than the theoretical value calculated in the
same ways as for the broken-TRI ensemble. It seems, therefore,
that the same distribution, Eq.~(\ref{fyod}), describes the
velocity distribution for both TRI case and the no-TRI situation.
The difference between the two ensembles (the former interpolating
between Poisson and GOE, the latter between Poisson and GUE)
appears in the numerical value of the velocity variance only. It is
clear the variance is a unique parameter that determines the
appropriate velocity scale, similarly as for GOE and GUE pure
ensembles \cite{Ga}.

\section{Distribution of level curvatures}

Let us consider now the distribution of curvatures, $K=d^2 E/d\lambda^2$.
As shown by Gaspard and coworkers \cite{Ga} the tail
of the distribution  decays algebraically
as $K^{-2-\beta}$. This universality has been
verified for different systems \cite{SHG91,TH92,ZD93}. At the same
time, the small
curvature behaviour has been found to be non-generic even
for strongly chaotic systems \cite{TH92,ZD93} and reflecting the
system-dependent wavefunction localization properties (scarring
by periodic orbits).

On the other hand, the scaled curvature
\begin{equation}
\kappa=K {\Delta \over \beta\pi \sigma_v^2},
\label{renorm}
\end{equation}
for pure random ensembles obeys
the generalized Cauchy distribution \cite{ZD93,Vo94,FS95}
\begin{equation}
P(\kappa)=N_\beta{1\over (1+\kappa^{2})^{\beta+2\over 2}},
\label{z-d}
\end{equation}
 (with $\beta=1,2,4$
for GOE, GUE, GSE, respectively, $N_\beta$ denotes the normalization constant).

Here we demonstrate that if one allows the  parameter
$\beta$ to acquire  real values, $\beta\in (0,2]$,
 the
same distribution may be used in a general case of the intermediate
ensemble interpolating between Poisson, GOE and GUE pure cases
(provided we consider the ``perpendicular'' transition).
 The normalization constant is then equal
\begin{equation}
N_\beta={1\over {\sqrt{\pi}}}
{\Gamma({\beta+2 \over 2}) \over \Gamma({\beta + 1\over 2})}.
\label{curvdistnorm}
\end{equation}
and the rescaling, Eq.~(\ref{renorm}) holds almost everywhere.

 To test this conjecture we have generated several parametric
dynamics "perpendicular" to crossovers between Poisson-GUE, Poisson--GOE and
GOE--GUE using, as before, the formulation of Section II,
 Eq.~(\ref{matrix1}) and Eq.~(\ref{pardyn}).
The numerically obtained histograms of curvatures in the double
logarithmic scale have been used to fit the algebraic decay of the
tail of the distribution to the formula $P(K)\sim K^{-\mu}$. Then
$\beta$ has been found as $\beta=\mu-2$ [compare Eq.~(\ref{z-d})].
The same value of $\beta$ has been used, together with the
numerically obtained velocity variance to rescale the curvatures
according to Eq.~(\ref{renorm}).  The exemplary
 results of such a procedure
together with the conjecture (\ref{z-d}) are presented
in Fig.~\ref{figu6} and Fig.~\ref{figu7} in double linear
and double logarithmic scale, respectively. Observe the
excellent agreement between the numerical results and the proposed
distribution.

While Eq.~(\ref{z-d}) seems to describe well, at least approximately,
the numerical data for curvatures everywhere in between pure
cases of GOE, GUE and Poisson limit, the scaling (\ref{renorm})
works best for the delocalized or weakly localized spectra. For
the Poisson -- GUE crossover, close to the Poisson limit,
the scaling obtained using Eq.(\ref{renorm}) is incorrect. The
agreement with the generalized Cauchy distribution (\ref{z-d})
 is obtained only if the numerical data are rescaled additionally by
a numerical factor of the order of unity (typically 1.5 -- 2, depending
on $\beta$). Putting it differently, the parameter
$\beta$ in the denominator of
Eq.(\ref{renorm}) should be replaced by other function of $\beta$
which goes to $\beta$ when transition to delocalized TRI broken
case (i.e. GUE) is fully accomplished. This indicates that
the proposed distribution
(\ref{z-d}) is most probably the approximate one only. Still we
find it quite remarkable that this simple analytic expression,
 with the proper rescaling, works so well for the
interpolating ensembles.

Let us mention that the power of the algebraic tail behaviour may be
analytically related to the level repulsion parameter $\beta$
by a simple consideration of $2 \times 2$ random matrices
\cite{Ga} yielding $\mu=\beta+2$. A comparison of $\beta$ values
obtained from the tail of the distribution with $\beta'$ values
obtained from the independent fit of the Izrailev distribution
 $P_{\beta'}(s)$ \cite{Iz90} is presented in Fig.~\ref{figu8} in
the whole interval of the intermediate $\beta$ values. The agreement
is quite good (and of the similar quality as that obtained for the
Fourier transform of the velocity-velocity correlation function)
considering that both the spacing distribution
$P_{\beta'}(s)$ and the proposed curvature distribution $P(\kappa)$
are most probably the good approximations to
the true distributions only.

It is worth noting that the distribution (\ref{z-d}) works well for
the ensemble interpolating between GOE and GUE for the perpendicular
action of the parameter $\lambda$. On the other hand, if $\lambda$
is responsible for the TRI symmetry breaking, it has been shown that
the tails of the curvature distribution are exponential \cite{KB94,BM94}
and not algebraic, as observed in this work. Parametric statistics are,
therefore, sensitive to
the way the parameter acts. Another example of this sensitivity
are available from  the earlier studies of periodic
band random matrices \cite{PBRM} and 3-D Anderson model \cite{ZIM94},
where the curvature distribution close to a log-norm distribution has been
observed in the localized case while yet another distribution
has been proposed in transition regime \cite{canal}.

\section{Velocity -- velocity correlation function}

In a series of papers, Altshuler and co--workers \cite{Alt} have
discussed the universality of parametric
statistical properties for disordered samples as well as for
Gaussian ensembles.
To reveal the universality both
the eigenvalues  (to unit mean spacing) and
the parameter $\lambda $ (as $X= \sigma_v \lambda$) have to be rescaled
\cite{Ga,ZD93,Alt}. We have observed the power of such a rescaling
already in the previous Sections.

Consider next
the velocity--velocity correlation function,
 a frequent subject of recent
investigations \cite{Alt,Be93,Za95,GZZ95,BLM96,SA93,KB94},

\begin{equation}
\label{corr}
C_v(\lambda ):=\frac 1{2\pi \Delta ^2}\bigl< \int\limits_0^{2\pi
}  v_i(\lambda^{\prime }) v_i(\lambda
^{\prime }+\lambda )  d\lambda^{\prime } \bigr>,
\end{equation}
where $\left\langle \text{ }\right\rangle $ denotes ensemble averaging
and $\Delta $ stands for the mean level spacing.
By definition $C_v(0)=\sigma_v^2$ thus
 the appropriately rescaled correlation functions take the form
$\tilde{C_v}(X)=C_v(X)/\sigma_v^2$.
Moreover, several models of time reversal
symmetry breaking due to Aharonov - Bohm flux lead to correlation
function practically indistinguishable from this characteristic for GUE.

It was shown  \cite{Alt,SA93,Be93}  that for all three  universality
classes the rescaled correlation function
\begin{equation}
 C_v(X)\sim A_\beta X^{-2},\qquad X\rightarrow\infty
\label{scalX}
\end{equation}
with the proportionality coefficient $A_\beta$ dependent on the
ensemble (we denote by $\beta$ the level repulsion parameter,
$\beta=1,2,4$ for GOE, GUE and GSE, respectively).

Explicit expressions have been obtained \cite{Alt} for a closely related
[but distinct from ${C_v(\lambda)}$] autocorrelation functions at
fixed energy.  A global approximation for ${\tilde{C_v}}(X)$ has been
proposed \cite{Za95}. For the case of a classically chaotic system
subject to
a Aharonov--Bohm flux Berry and Keating \cite{BK94} obtained a
semiclassical approximation for $C_v(\lambda)$ having the form of
an everywhere analytic function of $\lambda$.
Yet  it was demonstrated \cite{GZZ95}
that $C_v(\lambda)$ is not analytic  and suffers a logarithmic
singularity at $\lambda=0$.

 Analytic properties of correlation functions are conveniently
studied using the periodicity in $\lambda$.
In the Fourier domain
\begin{equation}
C_v(\lambda)=\sum_{n=0}^{\infty} c_n \cos(n \lambda).
\end{equation}
Mean squared velocity, determining the scale, is given by
the sum of all coefficients
$\sigma_v^2=C(0)=\sum\limits_{n=0}^{\infty }c_n.$
Expanding the dependence of a given  eigenvalue on $\lambda$ in the
Fourier series $E_i=\sum_{n=-\infty}^{\infty}a_ie^{i n\lambda}$,
where $a_n=a_{-n}^{*}$, on account of Eq.~(\ref{corr})
 it is easy to see that $c_n=n^2\langle|a_n|^2\rangle/\Delta^2$.

Asymptotic behavior of $C_v(\lambda)\sim -\lambda^{-2}$ corresponds to a linear
raise of Fourier coefficients $c_n$ for small $n$. On the other hand,
for large $n$ it was shown\cite{GZZ95} that
$\langle |a_n|^2\rangle \sim n^{-4-\beta}$ and consequently,
\begin{equation}
c_n\sim n^{-2-\beta}.
\label{scalcc}
\end{equation}
This result
 was obtained extending the parameter $\lambda$ into the complex
plane and analyzing the distribution of branch points and anticrossings
\cite{Wi89,ZK91}.

Thus, despite nonanalytic character of the correlation function
both $C_v(\lambda)$ and its Fourier transform have simple asymptotics
given by Eq.~(\ref{scalX}) and Eq.~(\ref{scalcc}), respectively. It is
interesting to see whether a similar behaviour may be found for
the interpolating ensembles. To this end we have studied the
asymptotics of $C_v$ for all three possible transitions, i.e.,
ensembles interpolating between Poison and GOE, Poisson and GUE,
as well as GOE and GUE. In all cases the parameter $\lambda$ acted
perpendicularly to a given transition (compare Fig.~\ref{scheme}).

We have observed the same $X^{-2}$ large
$X$ behaviour, Eq.~(\ref{scalX}) independently of the ensemble studied.
As an example
  Fig.~\ref{figu4} shows  the rescaled
  velocity-velocity correlation  functions
${\tilde{C_v}}(X)$ corresponding to 5 different cases along the
 Poisson-GUE crossover.

  Algebraic decay of the corresponding Fourier
transforms is visualized in Fig.~\ref{figu5}.
 Observe the continuous change of
the slope, growing from  $-4$ for GUE till $-2$
for the Poisson limit. This corresponds to the continuous change
of the repulsion parameter $\beta$ between 2 and 0 in the level
spacing distribution. Independently one may fit
the nearest neighbour spacing distribution obtained
numerically to
the Izrailev distribution\cite{Iz90}, $P_{\beta'}(s)$,
which provides an excellent approximation for the nearest neighbour
 spacing distribution
for  the interpolating ensembles. $P_{\beta'}(s)\sim s^{\beta'}$
for small $s$. We have verified that $\beta$ values obtained by fitting
the straight line to the tail of $\log(c_n)$ equal $\beta'$ values
obtained from the fits of the spacing distribution within 5\%.
Therefore, we conclude
that the validity of Eq.~(\ref{scalX}) and Eq.~(\ref{scalcc}) extends
to the intermediate ensembles and fractional values of the repulsion
parameter $\beta$.

We stress again that this result is restricted to the ``perpendicular''
transitions only. If parameter $\lambda$ is responsible for the
transition between the ensembles (e.g., the magnetic flux in the
Aharonov -- Bohm effect)
the velocity-velocity correlation function obeys the
$c_n\sim n^{-4}$ algebraic decay, independently of the
degree of localization \cite{GZZ95}.
Thus, similarly to the velocity distribution itself \cite{THSA} also
the velocity-velocity correlation function
${\tilde{C_v}}(X)$ is sensitive to  the nature of
perturbation generating the parametric dynamics.

In a full analogy with
the velocity-velocity correlation function (\ref{corr})
we define the curvature correlation function

\begin{equation}
\label{cork}
C_k(\lambda ):=\frac 1{2\pi \Delta ^2}\bigl< \int\limits_0^{2\pi
}  K_i(\lambda^{\prime }) K_i(\lambda
^{\prime }+\lambda )  d\lambda^{\prime } \bigr>.
\end{equation}

However, this function does not provide us with any new information.
This fact  is easy to understand studying the Fourier expansion
$C_k(\lambda)=\sum_{n=0}^{\infty}k_n e^{i n\lambda}$.
As for velocity correlation function one uses
 mean Fourier coefficients of individual energy levels
and obtains relation  $k_n=n^4 \langle |a_n|^2\rangle /\Delta^2$.
A comparison
with the velocity correlation function Fourier coefficients
yields immediately
\begin{equation}
C_k(\lambda)=\frac{\partial^2}{\partial\lambda^2}C_v(\lambda),
\label{nogood}
\end{equation}
which easily yields the properties of $C_k(\lambda)$ from known
properties (e.g. the asymptotic behavior) of $C_v(\lambda)$.
Eq.~(\ref{nogood}) holds for an arbitrary matrix ensemble. For
completeness we present the numerically obtained $C_k(\lambda)$,
 rescaled with respect to $C_k(0)$,
for GOE and GUE ensembles in Fig.~\ref{figu9}.
 Notice a much faster decay of the correlation between
curvatures as compared with the velocity correlation function.
Asymptotically, using $C_v\sim X^{-2}$ and Eq.~(\ref{nogood})
we get $C_k \propto X^{-4}$ for large rescaled parameter $X$, in
full agreement with Fig.~\ref{figu9}. The
curvature correlation function, in view of Eq.~(\ref{nogood}),
may be used, together with the velocity correlation function, for
numerical tests of the accuracy of curvature evaluation (which
may be quite tricky using the finite difference method since
 small and large curvatures may require different step in the
parameter).

\section{Higher derivatives of energy with respect to the parameter}

Algebraic decay of the Fourier transform of velocity
correlation function on one hand provides an
information about singularity of some higher derivative of
$C_v(\lambda)$ at $\lambda=0$, and on the other hand,
indicates a possibility of a divergence of a distribution variance
of the some  higher derivatives of
the energy levels with respect to the parameter\cite{GZZ95}. In
particular, for
orthogonal ensemble $(\beta=1)$, the variance of curvature distribution
$\langle K^2\rangle$ does not exist and in order to characterize the
mean curvature one uses the mean absolute value
$\langle|K|\rangle$ instead
\cite{PBRM}.  Moreover,
 the second moment  of the distribution of
third derivatives of energy levels $L:=d^3 E/d\lambda^3$
was predicted \cite{GZZ95} to diverge for
$\beta=1$ and $\beta=2$.

To test this prediction we have studied the distribution of these
third derivatives. The most difficult part here is to find how
to call them -- using the level motion picture where the curvature
of the level is identified with the acceleration of the fictitious
particle, the third derivative of the energy will correspond to
the derivative of the acceleration. In spirit of this mechanical
analogy we refer to the third derivative as a {\sl jerk} (editorial --
referee's -- reader's help here, whichever comes first, would be of great
value for the authors).

We have restricted  the numerical study of the distribution
of jerks to canonical orthogonal and unitary ensembles (GOE and
GUE).
The obtained numerical results are displayed in Fig.~\ref{figu10}. As
expected the distributions of jerks are characterized by the
algebraic tails; the numerically obtained power law decay yields
 $P(L)\sim L^{(\beta+3/2)}$. That  confirms the
divergence of the variances both for GOE and GUE.

It is interesting that the distribution of jerks may be quite nicely
approximated by a very simple ansatz
\begin{equation}
P(L)=\frac{{\cal N}'_\beta}{(1+B_\beta L^{A(\beta)})
^\frac{(\beta+3)}{2A(\beta)}},
\label{guess}
\end{equation}
where ${\cal N}'_\beta$ is a normalization constant while $B_\beta$
and $A(\beta)$ are free parameters.
In Eq.~(\ref{guess}) the jerks are conveniently rescaled taking the
unfolded spectrum (i.e. with the mean level spacing equal to unity)
with derivatives calculated with respect to the rescaled parameter,
$X=\sqrt{\langle v^2\rangle}\lambda$.

We have fitted the distribution of this form to the numerical
data for both GOE and GUE. The results are represented in Fig.~\ref{figu10}
as smooth curves and quite successfully represent the numerical data.
The obtained values of the parameters are equal for GOE to $A_1=1.67$
and $B_1=9.08$ while for GUE we obtain $A_2=2.50$ and $B_2=0.84$.
The obtained values of $A_\beta$ are close to simple fractions,
$A_1=5/3$ and $A_2=5/2$ - the corresponding curves are indistinguishable
from best fits within the accuracy of our data.

\section{Concluding remarks}

    We have analyzed various aspects of parametric dynamics in the
space of Hermitian random matrices. Such a model may be applied to
study transitions between Poissonian, orthogonal and unitary universality
classes. We have analyzed numerically the situations when the
parameter change does not modify the global properties of the
ensemble studied, the case baptized as a ``perpendicular''
transition to contrast it with the ``parallel'' case when
the parameter is responsible for the break up of the symmetry
or other change of the properties of the ensemble studied. We
have, however, compared our results with predictions of other
works where often such a ``parallel'' parameter action was
considered.

We have paid a particular attention to the study of the transition
between the Poissonian ensemble characterized by strongly
localized wavefunctions and the delocalized Gaussian ensembles (GOE
or GUE). In particular, the numerical tests of the analytic
predictions for the distribution of level velocities in the case
of broken time reversal invariance \cite{Fy94} confirmed the
predicted shape. We have observed a disagreement between
the theory \cite{Fy94} and the numerical data, however, as far
as the prediction for the velocity variance is considered. We have
discussed the possible origin of this difference. We have shown
that the distribution of the same functional form works
for the delocalization transition also for the real symmetric random
matrices.
 This calls for the extension of the theory to such
a case.

We have studied in detail also the distribution of level curvatures.
Here no analytic prediction is available. We have found that the
numerically obtained
distribution of curvatures is well approximated by the generalized Cauchy
distribution (\ref{z-d}) earlier shown to be exact
  \cite{ZD93,Vo94,FS95} for
the canonical, GOE and GUE ensembles. The only required modification
is to take the fractional value of the level repulsion parameter
$\beta$ in accordance with the spacing distribution. We have found
also that the same rescaling (\ref{renorm}) holds everywhere
except for the localized, no-TRI ensemble. Then the agreement
with Eq.~(\ref{z-d}) requires additional multiplication of all
curvatures (rescaling) by a factor of the order of unity and
dependent on $\beta$.

This form of the curvature distribution implies its algebraic tails
of the form $P(K)\sim K^{-2-\beta}$. Similarly we have found that
the same level repulsion parameter $\beta$ governs the tails
of the Fourier transformed
velocity correlation functions. Explicitly, the corresponding Fourier
coefficients satisfy to a good precision
$c_n\sim n^{-2-\beta}$.

Comparison with other works, where mostly the ``parallel''
transition have been studied \cite{THSA,KB94,BM94,IMZ96,PBRM,ZIM94}
on various models indicates strong differences with the
``parallel'' transitions.
This difference has been first observed
for the velocity correlation function in the case of a partially
broken TRI in a fully delocalized case \cite{THSA}.
Here we have shown that the sensitivity of level dynamics
 to the way in which the parameter acts extends also to other
parametric statistical measures as well as
to other ensembles interpolating between ``pure'' cases of
the Poisson, GOE and GUE ensembles. This has an important
consequence --- it shows that the universality of parametric
dynamics is more limited that anticipated before \cite{Alt}.

Finally consider the consequence of the presented results for
studies of realistic systems. Consider the semiclassical limit
when the system is ``large'', with a high density of states and
many highly excited levels. In the generic situation a small change
of the parameter cannot induce significant changes in system
symmetries and global properties - small changes of a parameter
may be thus considered as ``perpendicular'' cases. This indicates
that the ``perpendicular'' transitions studied in this paper are
typically generic. An important exception is the change of
the magnetic fields in systems with no additional symmetries where
the field induces the breakup of the time reversal symmetry acting,
therefore, in a ``parallel'' way.

Thus the fact that the curvature distribution for
the transition between the Poisson ensemble and GOE, realized via
the banded matrix model studied here, is described
by the generalized Cauchy law, Eq.~(\ref{z-d}) has important
consequences. It has been suggested (see \cite{ZS96} and references
cited there) that banded matrices may be used to simulate
statistical properties of partially chaotic systems interpolating
between the integrable case (with Poisson level spacing statistics)
and fully chaotic case (GOE). This has been partially based on the
similarity of the level spacing distribution observed in both
cases, well approximated by the Izrailev ansatz \cite{Iz90} or
for TRI systems by the Brody distribution.
However, even for chaotic systems,
as shown before, small curvature behaviour may be abundant due to
isolated avoided crossings and scarring of wavefunctions \cite{ZD93}.
For the mixed phase space systems the avoided crossings are typically
quite narrow and isolated -- between them the levels can be adiabatically
followed as a parameter is varied. Generically, small changes of
a parameter are accompanied by small changes of eigenenergies which
may be treated by a Taylor series expansion with a leading linear term.
It shows that such systems will exhibit a great abundance of
small curvatures, accompanied possibly by a singularity of the
distribution at $K=0$ (see also \cite{ZD93} for an additional discussion
and numerical examples). Banded random matrices show a
different curvature distribution than expected for a quantum system
with a mixed phase space. Thus this ensemble is not
adequate for simulating the statistical properties of partially
integrable or weakly chaotic systems at least in cases when
parametric dynamics is concerned. On the other hand, HBRM ensemble
seems to be very useful for obtaining predictions for
random systems which exhibit a transition to localization,
well into the localization regime.

\section{acknowledgements}

It is a pleasure to thank Italo Guarneri
and Yan Fyodorov
for fruitful   discussions. We acknowledge financial support
 by Polish Committee of
Scientific Research under  the Grant No.~2P03B~03810.


\begin{figure}
\caption{Scheme
 of the space of random matrices (and dynamical
systems). Three circles represent universality classes: Poissonian,
orthogonal and unitary, and dashed lines represent crossovers between
them. Solid arrows stand for "perpendicular" transitions, analyzed in
this paper. A broken arrow exemplifies a ``parallel'' transition, not
treated here.}
\label{scheme}
\end{figure}

\begin{figure}
\caption{Velocity distribution in a semilogarithmic scale for HBRM
model interpolating between Poisson ensemble and GUE ($\alpha=1$,
a fully broken TRI).
Thick (thin) line histogram corresponds to numerical data obtained in
 the localized case $x=0.126$ (the delocalized case, $x=1.408$) from
 diagonalizations of 10000 matrices of rank $N=71$.
  A dashed thick line represents the theoretical
prediction, Eq.~(\protect{\ref{fyod}}) while a thin line represents the
Gaussian distribution.
}
\label{figu2}
\end{figure}

\begin{figure}
\caption{Same as Fig.~2 but for real symmetric matrices, $\alpha=0$:
thick line histogram - a localized case, $x=0.056$, while thin line
histogram - a delocalized case, $x=5.63$.}
\label{figu3}
\end{figure}

\begin{figure}
\caption{Exemplary curvature distribution for the ensemble
interpolating between the localized Poisson case ($\beta=0$)
and GOE ($\beta=1$) obtained numerically for matrices of
rank $N=71$ and bandwidth $b=5$ corresponding to $x=0.352$ (histogram).
Thick dashed
line represents the fitted distribution, Eq.~(\protect{\ref{z-d}})
with $\beta=0.59$.
Thin solid and dotted lines represent the limiting distributions
for the Poisson ensemble and the GOE, respectively.
}
\label{figu6}
\end{figure}

\begin{figure}
\caption{Same as Fig.~\protect{\ref{figu6}} but in a double
logarithmic scale.}
\label{figu7}
\end{figure}

\begin{figure}
\caption{Parameter $\beta$ obtained from the decay of the tail of the curvature
distribution against the
  level repulsion parameter $\beta'$ obtained from the independent fit
  of Izrailev distribution \protect{\cite{Iz90}}. Each dot represents one
  ensemble interpolating between the Poisson ensemble, GOE or GUE.}
\label{figu8}
\end{figure}

\begin{figure}
\caption{Rescaled velocity correlation function
$\tilde{C_v}(X)$ obtained for 5
transitions perpendicular to the GUE-Poisson crossover:
 $N=71, c=2.0$;  $b=71 (\triangle)$ (GUE),  $b=10 (+)$,
 $b=7(\Box)$, $b=5 (\circ)$ and $b=4 (\Diamond)$.
Ensemble
averaging performed over $100$ matrices; the lines are drawn to guide
the eye. }
\label{figu4}
\end{figure}

\begin{figure}
\caption{Fourier transform of velocity correlation functions
displayed in Fig.~\protect{\ref{figu4}} in the log-log scale. Lines represent slopes
characteristic for GUE $(-4)$, GOE $(-3)$,  and Poisson $(-2)$.
 The data for
$b=10$ are not plotted to improve legibility of the figure.}
\label{figu5}
\end{figure}

\begin{figure}
\caption{Curvature--curvature  correlation function  ${\tilde{C_k}}(X)$
  obtained for $N=50$ $(\Diamond )$,
 $60 (\circ )$, $70 (\Box )$,
 $80 (\triangle )$, and $N=90 (\heartsuit /\clubsuit)$
 for GOE (open
symbols) and GUE (full symbols). Universal rescaled velocity
correlations are represented for comparison by thick dashed (GOE) and solid
(GUE) lines.    }
\label{figu9}
\end{figure}

\begin{figure}
\caption{Distribution of jerks (third derivatives of the eigenenergy with
respect to the parameter) $P(L)$
in the double logarithmic scale. Histograms correspond to numerically
obtained data for GOE (thick line) and GUE (thin line) both obtained for
random matrices of rank $N=71$. Dotted and dashed lines represent the best
fit of the proposed distribution, Eq.~(\protect{\ref{guess}}), for GOE and
GUE, respectively.
}
\label{figu10}
\end{figure}

\end{document}